\documentclass[usenatbib,fleqn]{mnras}
 \pdfoutput=1 
\usepackage{newtxtext,newtxmath}
\usepackage{algorithm,algorithmic}
\usepackage[T1]{fontenc}
\usepackage{ae,aecompl}

\usepackage[dvips]{graphicx}
\usepackage{epsfig,amsmath,color}
\newcommand{\beq}{\begin{equation}}
\newcommand{\eeq}{\end{equation}}
\newcommand{\beqn}{\begin{eqnarray}}
\newcommand{\eeqn}{\end{eqnarray}}

\def\bmath#1{\mbox{\boldmath$#1$}}

\long\def\symbolfootnote[#1]#2{\begingroup%
\def\thefootnote{\fnsymbol{footnote}}\footnote[#1]{#2}\endgroup}

\usepackage{pstricks,pst-text}
\floatname{algorithm}{Algorithm}

\title[Spatially constrained calibration]{Spatially constrained direction dependent calibration}
\author[Yatawatta]{Sarod Yatawatta$^{1}$\thanks{E-mail:
yatawatta@astron.nl}\\
$^{1}$ASTRON, Oude Hoogeveensedijk 4, 7991 PD Dwingeloo, The Netherlands}

\begin{document}
\date{\today}
\pagerange{\pageref{firstpage}--\pageref{lastpage}} \pubyear{2018}
\maketitle
\label{firstpage}
\begin{abstract}
Direction dependent calibration of widefield radio interferometers estimates the systematic errors along multiple directions in the sky. This is necessary because with most systematic errors that are caused by effects such as the ionosphere or the receiver beam shape, there is significant spatial variation. Fortunately, there is some deterministic behavior of these variations in most situations. We enforce this underlying smooth spatial behavior of systematic errors as an additional constraint onto spectrally constrained direction dependent calibration. Using both analysis and simulations, we show that this additional spatial constraint improves the performance of multi-frequency direction dependent calibration.
\end{abstract}
\section{Introduction}\label{sec:intro}
One limitation for improving the quality of data obtained by modern wide-field radio interferometric arrays is the systematic errors affecting the data. Calibration, or the determination and removal of systematic errors is therefore a crucial data processing step in radio interferometry. When the observing field of view is wide, especially in low-frequency radio astronomy, the systematic errors vary with the direction in the sky and therefore, direction dependent calibration is necessary. Numerous techniques exist for direction dependent calibration and \cite{Cotton2021} give a comprehensive overview of the current state-of-the-art.

By definition, 'direction dependence' implies that the systematic errors are spatially variable (in addition to temporal and spectral variability). We can categorize the spatial variation into two groups. In deterministic variability, the spatial variations are smooth and contiguous and they can be described by a simple model. Examples for such variations are the systematic errors caused by the main lobe of a phased array beam or the refraction caused by a benign ionosphere. The second category of spatial variability is random or stochastic. In this case, a simple model will not be able to completely describe the systematic errors. The side lobe patterns of a phased array beam or scintillation caused by a turbulent ionosphere are two examples for the causes of stochastic direction dependent errors. In practice, the systematic errors are a combination of both deterministic and stochastic components, and the amount of each component is unknown.

Multi-frequency direction dependent calibration has already been improved by using  the spectral smoothness of systematic errors as a regularizer \citep{DCAL,Brossard2016,DMUX,OLLIER2018,Y2020}. Improving on this, in this paper, we add an implicit model for the spatial dependence of systematic errors as an extra regularizer. We favor an implicit model over an explicit model because the amount of deterministic behavior of the direction dependence is unknown to us. For one observation, we might get well defined direction dependence of systematic errors while for another observation, we might get completely random and uncorrelated direction dependence. Our spatial model will adapt itself to each observation according to this dichotomy. We use the calibration solutions for each direction to bootstrap our spatial model. In order to prevent overfitting, we pass the solutions through an information bottleneck by using elastic net regression \citep{ElasticNet} to construct the spatial model.

Explicit models have already being used to improve direction dependent calibration, for example by modelling ionospheric effects \citep{Cotton,Intema,RTS,arora2,Albert2020} or beam effects \citep{B2008,DistModel,Cotton2021}. Most of these methods are not scalable to process multi-frequency data in a distributed computer. Instead of having a model with both spectral and spatial dependence, we use disjoint models and separate constraints for spectral dependence and spatial dependence. We use consensus optimization \citep{boyd2011} and extend our previous work \citep{Y2020} to incorporate the spatial constraint as an extension to a model derived by federated averaging \citep{McMahan2016}. In this manner, we are able to incorporate both spectral and spatial constraints without losing scalability in processing multi-frequency data in a distributed computer. However, since we use an implicit model, relating this model to actual physical phenomena such as the ionosphere and the beam shape is more involved and is a drawback of our approach.

The remainder of the paper is organized as follows. In section \ref{sec:model}, we define the radio interferometric data model and describe our spatially constrained distributed calibration algorithm. In section \ref{sec:perf}, we derive performance criteria based on the influence function \citep{Hampel86,ST2019}. We perform simulations and present the results in \ref{sec:simul} to illustrate the improvement due to the spatial constraint. Finally, we draw our conclusions in  section \ref{sec:conclusions}.

{\em Notation}: Lower case bold letters refer to column vectors (e.g. ${\bf y}$). Upper case bold letters refer to matrices (e.g. ${\bf C}$). Unless otherwise stated, all parameters are complex numbers. The set of complex numbers is given as ${\mathbb C}$ and the set of real numbers as  ${\mathbb R}$. The matrix inverse, pseudo-inverse, transpose, Hermitian transpose, and conjugation are referred to as $(\cdot)^{-1}$, $(\cdot)^{\dagger}$, $(\cdot)^{T}$, $(\cdot)^{H}$, $(\cdot)^{\star}$, respectively. The matrix Kronecker product is given by $\otimes$. The vectorized representation of a matrix is given by $\mathrm{vec}(\cdot)$. The identity matrix of size $N$ is given by ${\bf I}_N$. All logarithms are to the base $e$, unless stated otherwise. The Frobenius norm is given by $\|\cdot \|$. 

\section{Spectrally and spatially constrained calibration}\label{sec:model}
We consider a radio interferometer with $N$ receivers (or stations), collecting data from $K$ directions in the sky. The visibilities at baseline $p,q$ are given by \cite{HBS}
\beq \label{V}
{\bf V}_{pqf_i}=\sum_{k=1}^K {\bf J}_{pkf_i} {\bf C}_{pqkf_i} {\bf J}_{qkf_i}^H + {\bf N}_{pq}
\eeq
where ${\bf V}_{pqf_i}$ $(\in \mathbb{C}^{2\times 2})$ is the visibility matrix in full polarization. The subscripts $p$ and $q$ denote the two stations forming the baseline $p,q$ and $f_i$ denotes the receiving frequency. The systematic errors along direction $k$ are given by ${\bf J}_{pkf_i}$ and ${\bf J}_{qkf_i}$ $(\in \mathbb{C}^{2\times 2})$. The source coherency of direction $k$ is given by ${\bf C}_{pqkf_i}$ $(\in \mathbb{C}^{2\times 2})$. The noise term is given by ${\bf N}_{pq}$ $(\in \mathbb{C}^{2\times 2})$, which represents the receiver noise and the noise due to sources not included in the model. Note that all components in (\ref{V}) are also dependent on time but we implicitly assume this.

We augment the systematic errors along direction $k$ for all stations into a block matrix as
\beq
{{\bf J}_{kf_i}}\buildrel\triangle\over=[{\bf{J}}_{1k{f_i}}^T,{\bf{J}}_{2k{f_i}}^T,\ldots,{\bf{J}}_{Nk{f_i}}^T]^T,
\eeq
where ${{\bf J}_{kf_i}}$ $\in \mathbb{C}^{2N\times 2}$.

The spectral constraint on the the systematic errors is given by \citep{DCAL}
\beq \label{BZ}
{\bf J}_{kf_i}={\bf B}_{f_i} {\bf Z}_{k}
\eeq
where ${\bf B}_{f_i}$ ($\in \mathbb{R}^{2N\times 2FN}$) is a basis in frequency, evaluated at $f_i$ and  ${\bf Z}_{k}$ ($\in \mathbb{C}^{2FN\times 2}$) is a matrix that accumulates information over all frequencies (say $P$ frequencies) pertaining to direction $k$. Note that ${\bf Z}_k$ is therefore independent of frequency, i.e., a 'global' model. The number of basis functions describing the frequency dependence is given by $F$.

In \cite{Y2020}, we have introduced an additional constraint
\beq \label{ZZ}
{\bf Z}_{k} = \overline{{\bf Z}}_k
\eeq
where $\overline{{\bf Z}}_k$ ($\in \mathbb{C}^{2FN\times 2}$) is an external model describing the frequency dependence. For example, we can build ${\bf Z}_k$ in (\ref{BZ}) using data at only a subset of the full frequency range. Similarly, we can build $\overline{{\bf Z}}_k$ using another subset of frequencies. By using the constraint (\ref{ZZ}), we are able to reach consensus between these different frequency subsets. This is useful when we are computationally limited to simultaneously process all available frequencies.

In this paper, we build our external model $\overline{{\bf Z}}_k$ in (\ref{ZZ}) using the information available to all directions, i.e., by varying $k$. Let ${\bmath \Phi}_k$ be a basis function in space (such as spherical harmonics) evaluated along the direction $k$. We can describe $\overline{{\bf Z}}_k$ as
\beq \label{Zbar}
 \overline{{\bf Z}}_k = {\bf Z} {\bmath \Phi}_k
\eeq
where  ${\bmath \Phi}_k \in\mathbb{C}^{2G\times 2}$ and the global spatial model, ${\bf Z} \in \mathbb{C}^{2FN\times 2G}$. The number of spatial basis functions is given by $G$. Note that ${\bf Z}$ is independent of $k$, i.e., the direction in the sky, and is therefore global, both spectrally and spatially. 

The spatial basis  can be constructed by a vector of basis functions ${\bmath \phi}_k$ ($\in \mathbb{C}^{G\times 1}$) evaluated along direction $k$ as
\beq
{\bmath \Phi}_k ={\bf I}_2 \otimes {\bmath \phi}_k.
\eeq
Note that with $G=1$ we get (federated) averaging.

With the spectral and spatial constraints (\ref{BZ}) and (\ref{ZZ}), we formulate the calibration problem as

\beqn \label{conscalib}
\{{\bf {J}}_{kf_i},\ldots,{\bf {Z}}_k:\ \forall\ k,i\}=\underset{{\bf {J}}_{kf_i},\ldots,{\bf {Z}}_k}{\rm arg\ min} \sum_i g_{f_i}(\{{\bf J}_{kf_i}:\ \forall k\})\\\nonumber
{\rm subject\ to}\ \  {\bf {J}}_{kf_i}={\bf {B}}_{f_i} {\bf {Z}}_k,\ \ i\in[1,P],k\in[1,K]\\\nonumber
{\rm and}\ \ {\bf {Z}}_k=\overline{{\bf Z}}_k, \ \ k\in[1,K].
\eeqn
The unconstrained calibration cost function at frequency $f_i$ for all $K$ directions is given by $g_{f_i}(\{{\bf J}_{kf_i}:\ \forall k\})$ and this can be solved for instance by using the space alternating generalized expectation maximization (SAGE) algorithm \cite{Fess94,Kaz2}.

In order to solve (\ref{conscalib}), we form the augmented Lagrangian, for all $k,i$ as
\beqn \label{aug}
\lefteqn{
L(\{{\bf {J}}_{kf_i},{\bf {Z}}_k,{\bf {Y}}_{kf_i},{\bf X}_{k}: \forall\ k,i\}) }\\\nonumber
&& =\sum_i g_{f_i}(\{{\bf J}_{kf_i}:\ \forall k\})\\\nonumber
&& + \sum_{i,k} \left( \| {\bf {Y}}_{kf_i}^H ({\bf {J}}_{kf_i}- {\bf {B}}_{f_i} {\bf {Z}}_k)\| + \frac{\rho}{2} \| {\bf {J}}_{kf_i}- {\bf {B}}_{f_i} {\bf {Z}}_k \|^2 \right) \\\nonumber
&&+\sum_k \left(\|{\bf X}_k^H\left(  {\bf {Z}}_k - \overline{{\bf Z}}_k \right)\|+ \frac{\alpha}{2} \|  {\bf {Z}}_k - \overline{{\bf Z}}_k \|^2 \right).
\eeqn
The Lagrange multiplier for the spectral constraint (\ref{BZ}) is given by ${\bf {Y}}_{kf_i}$ ($\in \mathbb{C}^{2N\times 2}$) while the Lagrange multiplier for the spatial constraint (\ref{ZZ}) is given by ${\bf X}_k$ ($\in \mathbb{C}^{2FN\times 2}$). The regularization factors for both constraints are given by $\rho$ and $\alpha$. Note that both $\rho$ and $\alpha$ can be made direction ($k$) dependent as well, but we omit this to simplify the notation.

We use the alternating direction method of multipliers (ADMM) algorithm \citep{boyd2011} to solve (\ref{aug}) and the pseudo-code is given in algorithm \ref{algSADMM}.
As in \citep{DCAL}, we consider a fusion centre that is connected to a set of worker nodes to implement algorithm \ref{algSADMM}. Most of the computations are done in parallel at all the worker nodes. 

\begin{algorithm}
\caption{Spatially constrained distributed calibration}
\label{algSADMM}
\begin{algorithmic}[1]
\REQUIRE $A$: number of ADMM iterations, $C$: cadence of spatial model update, $\rho,\alpha$: regularization factors
\STATE Initialize ${\bf {Y}}_{kf_i}$,$\overline{{\bf Z}}_k$,${\bf X}_k$ to zero $\forall k,i$
\FOR{$a=1,\ldots,A$}
\STATE \COMMENT{Do in parallel $\forall k,i$ at all compute agents and the fusion centre}
 \STATE Compute agents solve (\ref{aug}) for ${\bf {J}}_{kf_i}$
  \STATE Fusion centre solves (\ref{zsol}) for ${\bf Z}_k$
 \STATE ${\bf {Y}}_{kf_i} \leftarrow {\bf {Y}}_{kf_i} + \rho \left({\bf {J}}_{kf_i}-{\bf {B}}_{f_i} {\bf {Z}}_k\right)$
  \IF{$a$ is a multiple of $C$}
  \STATE Fusion centre updates ${\bf Z}$ using (\ref{Enet})
  \STATE Fusion centre updates $\overline{{\bf Z}}_k$ using (\ref{Zbar})
\STATE ${\bf X}_k \leftarrow {\bf X}_k + \alpha \left(  {\bf {Z}}_k - \overline{{\bf Z}}_k\right)$
 \ENDIF
\ENDFOR
\end{algorithmic}
\end{algorithm}

The fusion centre performs the update of ${\bf Z}_k$ in (\ref{BZ}) and also the update of ${\bf Z}$ in (\ref{Zbar}). Note that the spatial model ${\bf Z}$ is updated less-frequently than the spectral model ${\bf Z}_k$, i.e., with a cadence of $C$ in algorithm \ref{algSADMM}. The spectral model can be updated in closed form by solving

\beqn \label{zsol}
\lefteqn{{\bf {Z}}_k= }\\\nonumber
&& \left( \sum_i \rho {\bf {B}}_{f_i}^T {\bf {B}}_{f_i} + \alpha {\bf I}_{2FN} \right)^{\dagger}
\left(\sum_i {\bf {B}}_{f_i}^T \left({\bf {Y}}_{kf_i} + \rho {\bf {J}}_{kf_i}\right) + \alpha  \overline{{\bf Z}}_k -{\bf X}_{k} \right).
\eeqn

The spatial model update is performed using (\ref{ZZ}) and (\ref{Zbar}). In order to prevent overfitting, we use elastic-net regression \citep{ElasticNet} to update ${\bf Z}$. The model update can be formulated as
\beq \label{Enet}
{\bf Z} = \underset{{\bf Z}}{\rm arg\ min} \sum_k \|\overline{{\bf Z}}_k - {\bf Z} {\bmath \Phi}_k \|^2 + \lambda \|{\bf Z}\|^2 + \mu \|{\bf Z}\|_1
\eeq
where $\lambda$ and $\mu$ are introduced to keep ${\bf Z}$ low-energy and sparse. We use the fast iterative shrinkage and thresholding algorithm (FISTA \cite{FISTA}) to solve (\ref{Enet}) using the differentiable cost function
\beq \label{hh}
h({\bf Z})=\sum_k \|\overline{{\bf Z}}_k - {\bf Z} {\bmath \Phi}_k \|^2 + \lambda \|{\bf Z}\|^2
\eeq
and its gradient
\beq \label{gradf}
\nabla h = {\bf Z}\left( \sum_k {\bmath \Phi}_k {{\bmath \Phi}_k}^H + \lambda {\bf I}_{2G} \right) - \sum_k \overline{{\bf Z}}_k {{\bmath \Phi}_k}^H.
\eeq

In FISTA, we iterate with starting step size $t$, and at each iteration, we perform soft thresholding of each element of ${\bf Z}$ (real and imaginary parts taken separately) as 
\beq
\mathcal{T}([{\bf Z}]_i)={\rm sign}\left( [{\bf Z}-t \nabla h]_{i} \right) \left( |[{\bf Z}-t \nabla h]_{i}| - \mu t\right)_{+}
\eeq
 where we use the subscript $i$ to denote a single element of ${\bf Z}$, with real and imaginary parts taken separately. The step size $t$ is automatically updated in FISTA. The soft thresholding operator is denoted as $\mathcal{T}(\cdot)$ and $(z)_{+}={\rm max}(0,z)$.

It is evident from the construction of ${\bf Z}_k$ and ${\bf Z}$ that we do not build an explicit model for physical effects such as the beam shape or the ionosphere. Our model is implicit but much simpler because we decouple the spectral dependence and the spatial dependence. If we want to estimate the systematic errors along direction $k$ at frequency $f_i$ (subject to a unitary ambiguity), we can use (\ref{BZ}), (\ref{ZZ}) and (\ref{Zbar}) to get
\beq \label{3fac}
\widehat{\bf J}_{kf_i}={\bf B}_{f_i} {\bf Z} {\bmath \Phi}_k
\eeq
illustrating the decoupling of spectral and spatial coordinates. Another way of looking at (\ref{3fac}) is that we factorize the systematic errors along direction $k$ at frequency $f_i$ into three matrices (similar to a low-rank matrix approximation \cite{Lu2015}), out of which, we have some freedom in selecting ${\bf B}_{f_i}$ and ${\bmath \Phi}_k$. This emphasizes the careful selection of basis functions in frequency and in space.

\section{Performance analysis}\label{sec:perf}
We reuse our previous work \citep{SAM2018,SP2019} and derive the influence function  \citep{Hampel86} as a performance measure.
From (\ref{zsol}), we select a single frequency, i.e., $f_i=f$, and a single direction $k$. At convergence, the global spectral model ${\bf Z}_k$ can be written as a function of variables ${\bf J}_{kf}$ and ${\bf Y}_{kf}$ as 
\beq \label{glZk}
{\bf Z}_k=\rho {\bf P}{\bf J}_{kf} + {\bf P}{\bf Y}_{kf}+{\bf R}
\eeq
where
\beq
{\bf P}\buildrel \triangle \over = \left(\sum_i \rho {\bf {B}}_{f_i}^T {\bf {B}}_{f_i} + \alpha {\bf I}_{2FN} \right)^{\dagger}  {\bf B}_f^T\ \ \in  \mathbb{C}^{2FN\times 2N}
\eeq
and the remainder term ${\bf R}$ is independent of ${\bf J}_{kf}$ and ${\bf Y}_{kf}$. We substitute (\ref{glZk}) to (\ref{aug}) and find the gradient with respect to ${\bf J}_{kf}$ and ${\bf Y}_{kf}$. Let ${\bf F}\buildrel \triangle \over={\bf I}_{2N}-\rho{\bf B}_f{\bf P}$  ($\in \mathbb{C}^{2N\times 2N}$) and let ${\bf r}_1({\bf R}_1)$ and ${\bf r}_2({\bf R}_2)$ be the remainder terms that are independent of ${\bf J}_{kf}$ and ${\bf Y}_{kf}$. We can simplify the gradient of (\ref{aug}) as
\beqn \label{gradJ}
\lefteqn{{\rm grad}(L,{\bf J}_{kf})}\\\nonumber
&&={\rm grad}(g_{f}({\bf J}_{kf}),{\bf J}_{kf}) + \left(\frac{\rho}{2}{\bf F}^H{\bf F}+\frac{\alpha}{2}\rho^2 {\bf P}^H{\bf P}\right){\bf J}_{kf}\\\nonumber
&&+ \left(\frac{1}{2}{\bf F}^H{\bf F}+\frac{\alpha}{2}\rho{\bf P}^H{\bf P}\right){\bf Y}_{kf}+{\bf r}_1({\bf R}_1),
\eeqn
and
\beqn \label{gradY}
\lefteqn{{\rm grad}(L,{\bf Y}_{kf})}\\\nonumber
&&=\left(\frac{1}{2}{\bf F}^H{\bf F}+\frac{\alpha}{2}\rho {\bf P}^H{\bf P}\right){\bf J}_{kf}\\\nonumber
&&+ \left(-\frac{1}{2\rho}({\bf I}-{\bf F}^H{\bf F})+\frac{\alpha}{2}{\bf P}^H{\bf P}\right){\bf Y}_{kf}+ {\bf r}_2({\bf R}_2).
\eeqn

At a local minimum, we have ${\rm grad}(L,{\bf J}_{kf})={\bf 0}$ and ${\rm grad}(L,{\bf Y}_{kf})={\bf 0}$. Therefore, we can equate (\ref{gradJ}) and (\ref{gradY}) to zero to get a system of equations as
\beqn \label{nodiff}
\lefteqn{{\bf H}\left[ \begin{array}{c}
  {\bf J}_{kf}\\
  {\bf Y}_{kf}
\end{array} \right]}\\\nonumber
&&+
\left[ \begin{array}{c}
  {\rm grad}(g_{f}({\bf J}_f),{\bf J}_f)+{\bf r}_1({\bf R}_1)\\
  {\bf r}_2({\bf R}_2)\\
\end{array} \right] 
=
\left[ \begin{array}{c}
  {\bf 0}\\
  {\bf 0}\\
\end{array} \right] 
\eeqn
where 
\beq
{\bf H}=\left[ \begin{array}{cc}
  H_{11} & H_{12}\\
  H_{21} & H_{22}\\
\end{array} \right]
\eeq
with
\beqn
{\bf H}_{11}\buildrel \triangle \over =\left(\frac{\rho}{2}{\bf F}^H{\bf F}+\frac{\alpha}{2}\rho^2 {\bf P}^H{\bf P}\right)\\
{\bf H}_{12}={\bf H}_{21}^H\buildrel \triangle \over =\left(\frac{1}{2}{\bf F}^H{\bf F}+\frac{\alpha}{2}\rho{\bf P}^H{\bf P}\right)\\
{\bf H}_{22}\buildrel \triangle \over =\left(-\frac{1}{2\rho}({\bf I}-{\bf F}^H{\bf F})+\frac{\alpha}{2}{\bf P}^H{\bf P}\right).
\eeqn

Following \cite{SAM2018}, we take the differential of (\ref{nodiff}) to get
\beq \label{diff}
{\bf H} \left[ \begin{array}{c}
  d{\bf J}_{kf}\\
  d{\bf Y}_{kf}\\
\end{array} \right]
+
\left[ \begin{array}{c}
  d{\rm grad}(g_{f}({\bf J}_{kf}),{\bf J}_{kf})\\
  {\bf 0}\\
\end{array} \right] 
=
\left[ \begin{array}{c}
  {\bf 0}\\
  {\bf 0}\\
\end{array} \right] 
\eeq
where terms independent of ${\bf J}_{kf}$ and ${\bf Y}_{kf}$ become zero in the differential. By row elimination, we can simplify (\ref{diff}) to eliminate ${\bf Y}_{kf}$ to get
\beq \label{gradJ0}
 \widetilde{\bf H}d{\bf J}_{kf} + d{\rm grad}(g_{f}({\bf J}_{kf}),{\bf J}_{kf}) ={\bf 0}
\eeq
where 
\beq
\widetilde{\bf H} = {\bf H}_{11} - {\bf H}_{12}{\bf H}_{22}^{\dagger}{\bf H}_{21},\ \  \in \mathbb{C}^{2N\times 2N}.
\eeq
Using the influence function, we study the change in ${\bf J}_{kf}$, i.e., the solution to (\ref{conscalib}), due to small changes in the input data. For any given baseline $p,q$ at any given time and frequency, there are 8 real data points because ${\bf V}_{pqf_i}$ in (\ref{V}) is a matrix in $\mathbb{C}^{2\times 2}$ requiring 8 real data points. We select one baseline (say $p^\prime,q^\prime$) and one real valued data point out of 8 (say $r$). Let this data point be $x_{p^\prime q^\prime r}$. We apply the chain-rule to differentiate $g_{f}({\bf J}_{kf})$, and considering this to be a function of both ${\bf J}_{kf}$ and input data $x_{p^\prime q^\prime r}$,
\beqn \label{dgf}
d \mathrm{vec}\left({\rm grad}(g_{f}({\bf J}_{kf}),{\bf J}_{kf})\right) = \mathcal{D}_{\bf J}{\rm grad}(g_{f}({\bf J}_{kf})) \mathrm{vec}\left(d{\bf J}_{kf}\right)\\\nonumber
 +\frac{\partial}{\partial x_{p^\prime q^\prime r}} \mathrm{vec}\left({\rm grad}(g_{f}({\bf J}_{kf}),{\bf J}_{kf}) \right).
\eeqn
where $\mathcal{D}_{\bf J}{\rm grad}(g_{f}({\bf J}_{kf}))$ is the Hessian of $g_{f}({\bf J}_{kf})$ whose exact expression can be found in \cite{SP2019}. 
Substituting (\ref{dgf}) into (\ref{gradJ0}), we get
\beqn
 {\bf I}_2\otimes \widetilde{\bf H}\mathrm{vec}\left(d{\bf J}_{kf}\right) +\left(
\mathcal{D}_{\bf J}{\rm grad}(g_{f}({\bf J}_{kf})) \mathrm{vec}\left(d{\bf J}_{kf}\right) \right.\\\nonumber
+\left. \frac{\partial}{\partial x_{p^\prime q^\prime r}} \mathrm{vec}\left({\rm grad}(g_{f}({\bf J}_{kf}),{\bf J}_{kf}) \right) \right) = {\bf 0}.
\eeqn

Using the fact that only ${\bf V}_{p^\prime q^\prime f}$ in (\ref{V}) is dependent on $x_{p^\prime q^\prime r}$, we have
\beqn
 \lefteqn{\frac{\partial}{\partial x_{p^\prime q^\prime r}} \mathrm{vec}\left({\rm grad}(g_{f}({\bf J}_{kf}),{\bf J}_{kf}) \right)}\\\nonumber
&& = - \left({\bf A}_{q^\prime}{\bf J}_{kf} {\bf C}_{p^\prime q^\prime f}^H\right)^T \otimes {\bf A}_{p^\prime}^T \mathrm{vec}\left(\frac{\partial {\bf V}_{p^\prime q^\prime f}}{\partial x_{p^\prime q^\prime r}}\right)
\eeqn
where the canonical selection matrix ${\bf A}_p$ ($\in \mathbb{R}^{2\times 2N}$) is given as
\beq \label{Ap}
{\bf A}_p \buildrel\triangle\over=[{\bf 0},{\bf 0},\ldots,{\bf I}_2,\ldots,{\bf 0}].
\eeq
In other words, only the $p$-th block of (\ref{Ap}) is ${\bf I}_2$ and the rest are all zeros.

Finally we have
\beqn
\label{Jderiv}
\lefteqn{\mathrm{vec}\left(\frac{\partial {\bf J}_{kf}}{\partial x_{p^\prime q^\prime r}}\right)}\\\nonumber
&&=\left( \mathcal{D}_{\bf J}{\rm grad}(g_{f}({\bf J}_{kf}))+{\bf I}_2\otimes \widetilde{\bf H}\right)^{\dagger}\\\nonumber
&&\times \left({\bf A}_{q^\prime}{\bf J}_{kf} {\bf C}_{p^\prime q^\prime f}^H\right)^T \otimes {\bf A}_{p^\prime}^T \mathrm{vec}\left(\frac{\partial {\bf V}_{p^\prime q^\prime f}}{\partial x_{p^\prime q^\prime r}}\right)
\eeqn
that relates the change in ${\bf J}_{kf}$ due to a small change in the input datum $x_{p^\prime q^\prime r}$.

We use (\ref{Jderiv}) similar to \citep{Y2021} section 2.2 to study the influence function of the residual ${\bf R}_{pqf}$,
\beq \label{res}
{\bf R}_{pqf}={\bf V}_{pqf}-\sum_{k=1}^{K} \widehat{{\bf J}}_{kpf} {\bf C}_{kpqf} \widehat{{\bf J}}_{kqf}^H
\eeq
where $\widehat{{\bf J}}_{kpf} \forall k,p,f$ are calculated using the solution of (\ref{conscalib}).
We consider a mapping between the input data and the output residual as
\beq \label{resvec}
{\bf y}={\bf x}-{\bf s}(\widehat{{\bmath \theta}}).
\eeq
where ${\bf x}$ is the input and ${\bf y}$ is the residual, both $\in \mathbb{R}^D$, where the data length $D=8 \times N(N-1)/2 \times \Delta_t$ indicates the amount of data used to obtain a solution for (\ref{conscalib}), for one frequency. The model ${\bf s}(\cdot)$ describes the sky model and is parameterized by $\widehat{{\bf J}}_{kpf} \forall k,p,f$ which is represented as $\widehat{{\bmath \theta}}$ real parameters.

We use the statistical relation between the input ${\bf x}$ and the output ${\bf y}$by representing their probability density functions, $p_X({\bf x})$ and $p_Y({\bf y})$ as
\beq \label{pdf}
p_X({\bf x}) = |\mathcal{J}| p_Y({\bf y})
\eeq
where $\mathcal{J}$ is the Jacobian of the mapping between the input and the output, which is dependent on the model ${\bf s}(\widehat{{\bmath \theta}})$ in (\ref{resvec}) which in turn is dependent on (\ref{Jderiv}). The exact expressions can be found in \cite{SAM2018,SP2019}. The determinant of $\mathcal{J}$ is given as
\beq
|\mathcal{J}|=\exp \left( \sum_{i=1}^D \log(1 + \lambda_i) \right)
\eeq
where $\lambda_i$ are eigenvalues that are dependent on the model and the loss function (see equation 14 in \cite{Y2021}). Ideally, all $\lambda_i \rightarrow 0$ so we have $|\mathcal{J}|=1$, i.e., no distortion in the the probability density relation (\ref{pdf}). In practice, some $\lambda_i$ are non zero (and negative). Therefore, there is always a distortion due to the calibration. Another way of looking at this distortion is to measure the number of degrees of freedom absorbed into the calibration. This can be done by measuring the area of the epigraph of the curve $1+\lambda_i$ up to the ordinate 1. As an example, a linear system is considered in section 3.1 of \cite{SP2019}. We will give an example related to calibration in section \ref{sec:simul}.

\section{Simulations}\label{sec:simul}
We simulate an interferometric array with $N=62$ stations, using phased array beams such as LOFAR. The duration of the observation is 10 min, divided into samples with 10 sec integration, so in total $60$ time samples. We use $P=8$ frequencies, equally spaced in the range $[115,185]$ MHz. The calibration is performed using every $\Delta_t=10$ time samples, or 100 sec of data.

The sky model consists of $K=10$ point-sources in the sky that are being calibrated, with their positions randomly chosen within a $30$ deg radius from the field centre. Their intrinsic fluxes are randomly chosen in the range $[100,1000]$ Jy and their spectral indices are randomly chosen from a standard normal distribution. An additional $400$ weak sources (both point sources and Gaussians) are randomly positioned across a $16\times 16$ square degrees field of view. Their intensities are uniform-randomly selected from $[0.01,0.5]$ Jy with flat spectra. All the aforementioned sources are unpolarized. A model for diffuse structure (with Stokes I, Q and U fluxes) based on shapelets are also added to the simulation. The simulation incorporates beam effects, both the dipole beam and the station beam.

The systematic errors, i.e., ${\bf J}_{pkf_i}$ in (\ref{V}), are simulated for the $K=10$ sources as follows. For any given $p$, we simulate the 8 values of ${\bf J}_{pkf_i}$ for the central frequency and for $k=1$. We multiply this with a random 3-rd order polynomial in frequency to get the frequency variation. We also multiply this value with a random sinusoidal in time to get time variability. We get spatial variability by propagating ${\bf J}_{pkf_i}$ for $k=1$ to other directions (or other values of $k$). We do this by generating random planes in $l,m$ coordinates (such as $a_1 l+a_2 m+a_3$ where $a_1,a_2,a_3$ are generated from a standard normal distribution) and multiplying the 8 values of ${\bf J}_{pkf_i}$ for $k=1$ with these random planes evaluated at the $l,m$ coordinates of each direction. Finally, an additional random value drawn from a standard normal distribution is added to the values of ${\bf J}_{pkf_i}$ at each $k$.

\begin{figure*}
\begin{minipage}{0.98\linewidth}
\begin{center}
\begin{minipage}{0.48\linewidth}
\centering
 \centerline{\epsfig{figure=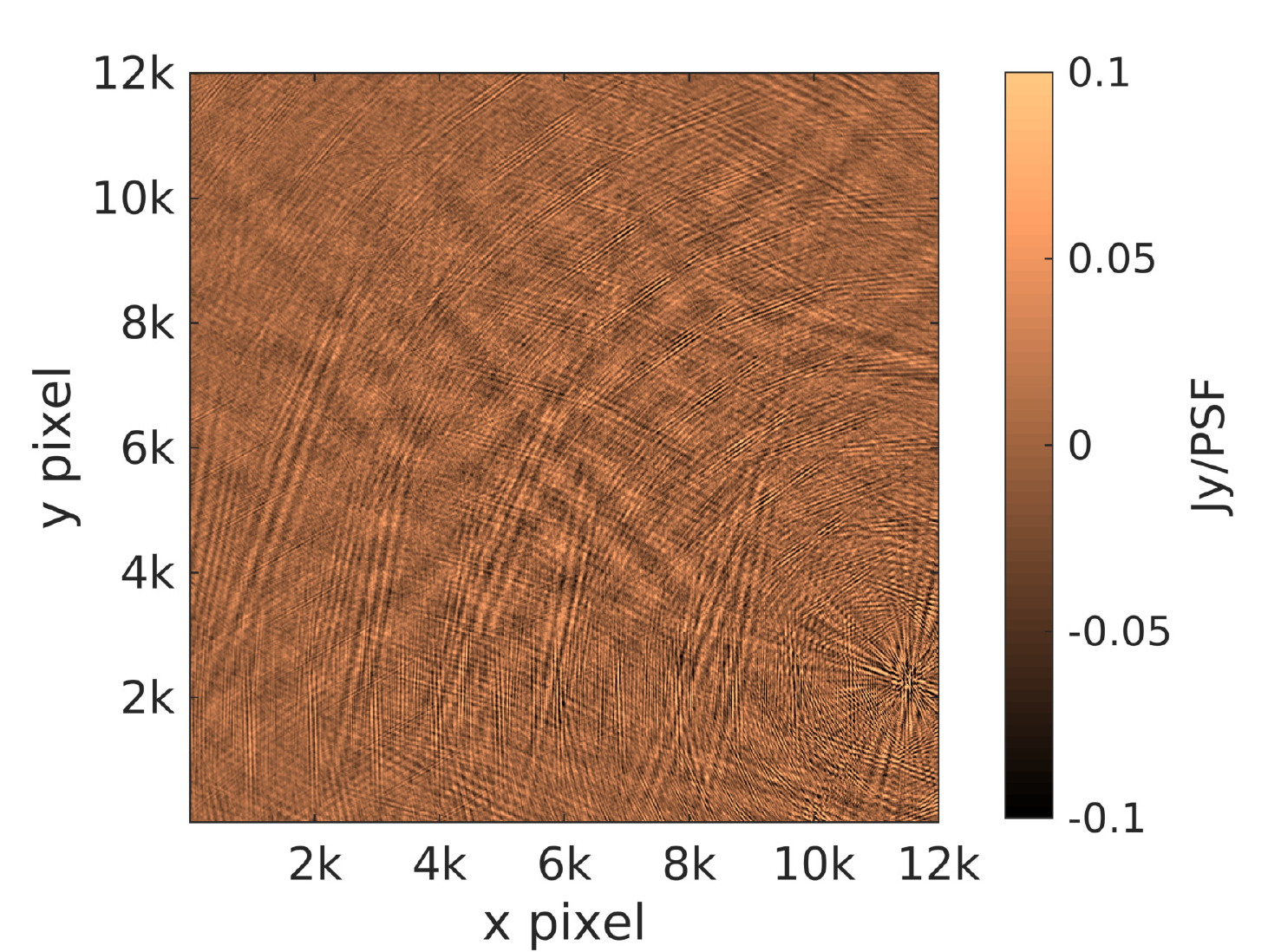,width=8.0cm}}
\vspace{0.5cm} \centerline{(a)}\smallskip
\end{minipage}
\begin{minipage}{0.48\linewidth}
\centering
 \centerline{\epsfig{figure=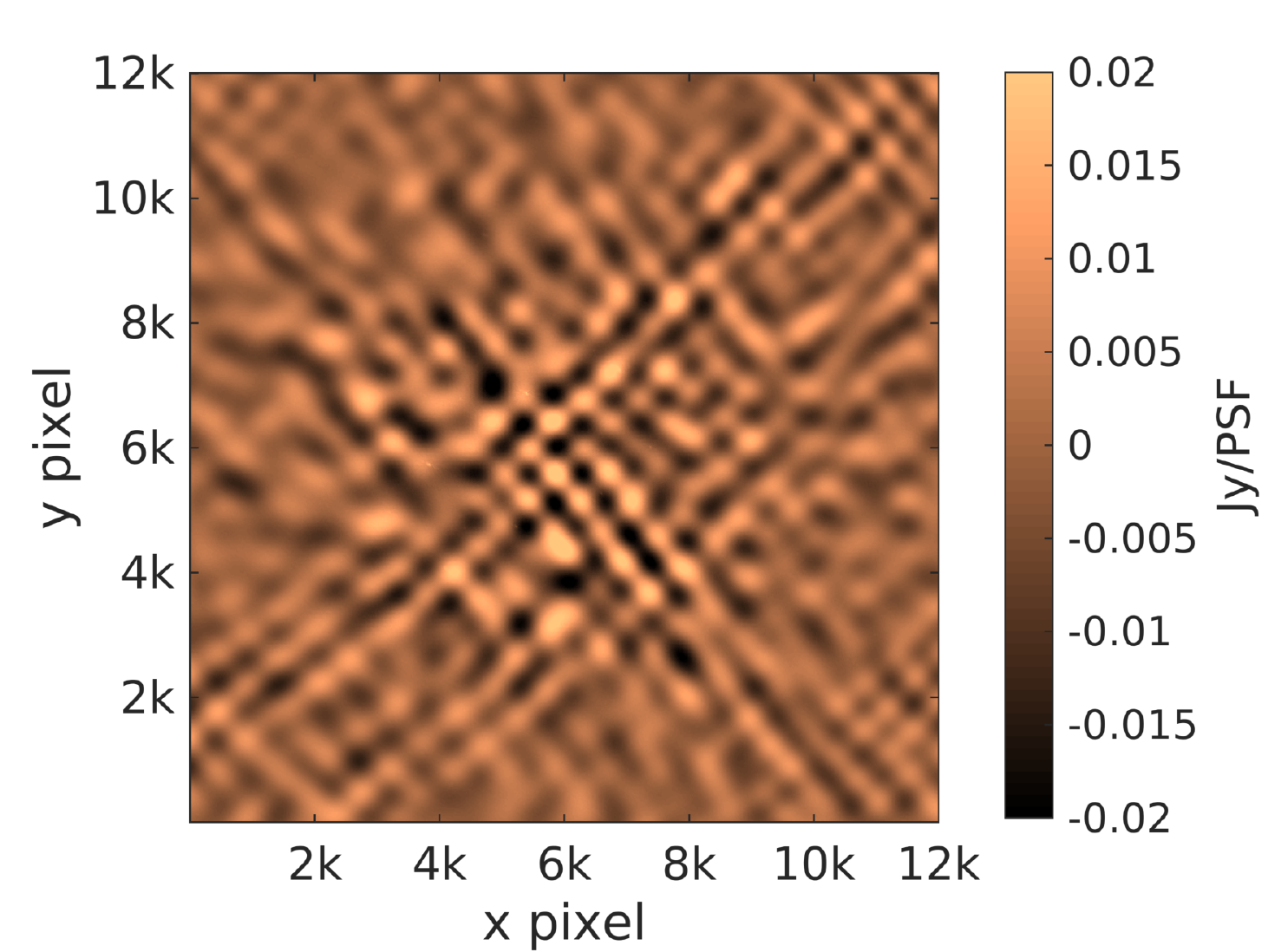,width=8.0cm}}
\vspace{0.5cm} \centerline{(b)}\smallskip
\end{minipage}\\
\begin{minipage}{0.48\linewidth}
\centering
 \centerline{\epsfig{figure=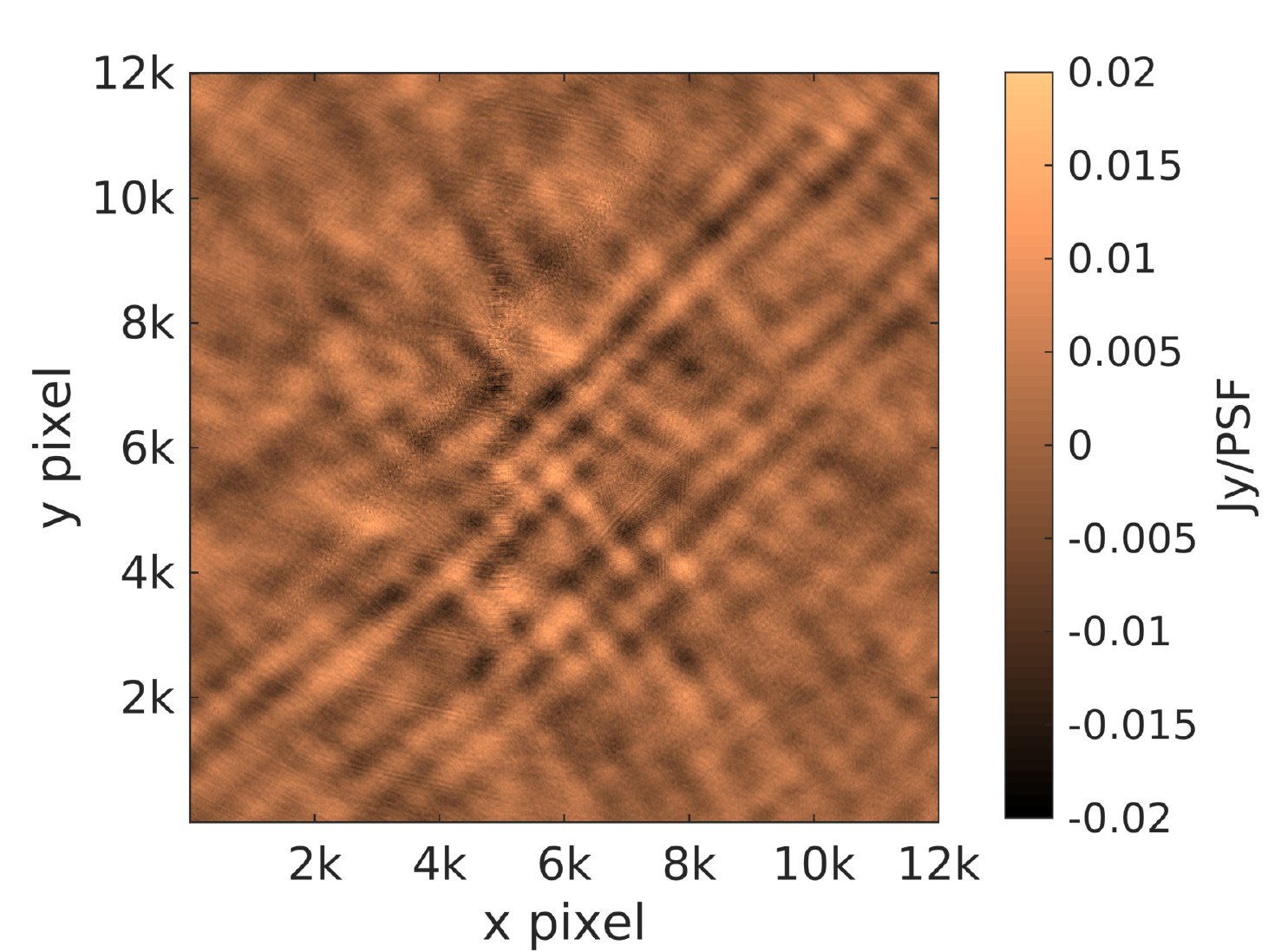,width=8.0cm}}
\vspace{0.5cm} \centerline{(c)}\smallskip
\end{minipage}
\begin{minipage}{0.48\linewidth}
\centering
 \centerline{\epsfig{figure=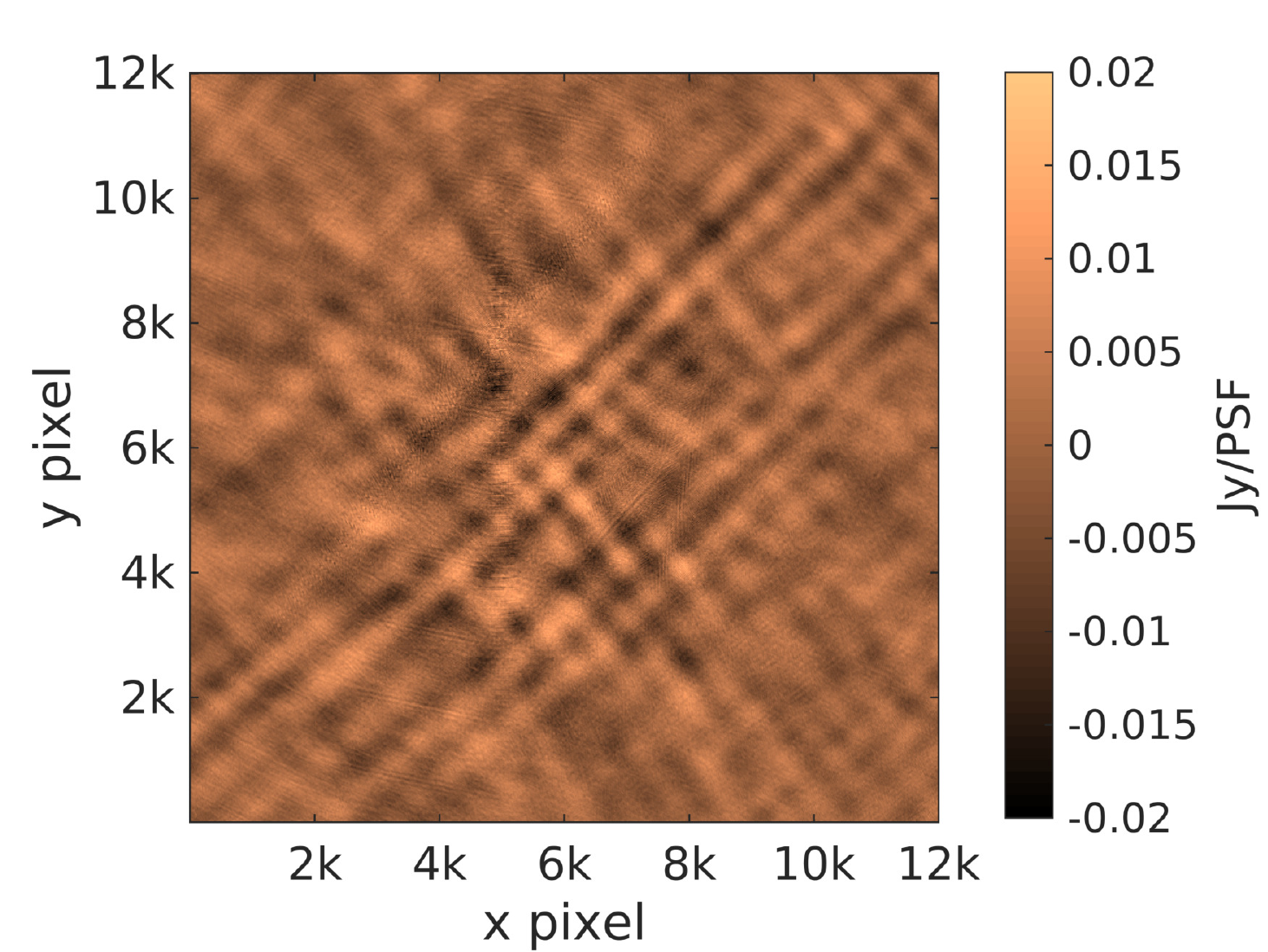,width=8.0cm}}
\vspace{0.5cm} \centerline{(d)}\smallskip
\end{minipage}
\end{center}
\caption{
  Sample images (not deconvolved) made using simulated data, covering about $13.3 \times 13.3$ square degrees in the sky. (a) The image before calibration (b) The diffuse sky and the weak sources that are hidden in the simulated data. (c) The residual image after calibration, without using spatial regularization. (d) The residual image after calibration, with spatial regularization. Both (c) and (d) reveal the weak sky shown in (b), but the intensity is much less.
\label{simulated_maps}}
\end{minipage}
\end{figure*}

The noise ${\bf N}_{pq}$ in (\ref{V}) is generated as complex zero mean Gaussian distributed elements and is added to the simulated signal with a signal to noise ratio of $0.05$. An example of a simulation is shown in Fig. \ref{simulated_maps} (a). 

We perform calibration using a 3-rd order Bernstein polynomial in frequency for consensus. The spectral regularization parameter $\rho$ is chosen with a peak value of $300$ and scaled according to the apparent flux of each of the $K=10$ directions being calibrated. When spatial regularization is enabled, we use a spherical harmonic basis with order $3$ (giving $G=9$ basis functions). The spatial regularization parameter $\alpha$ is set equal to $\rho$ for each direction. We use 20 ADMM iterations and the spatial model is updated at every 10-th iteration (if enabled). The elastic net regression to update the spatial model uses $\lambda=0.01$ and $\mu=10^{-4}$ as the L2 and L1 constraints and we use $40$ FISTA iterations.

\begin{figure}
\begin{minipage}{0.98\linewidth}
\begin{center}
\centering
\centerline{\epsfig{figure=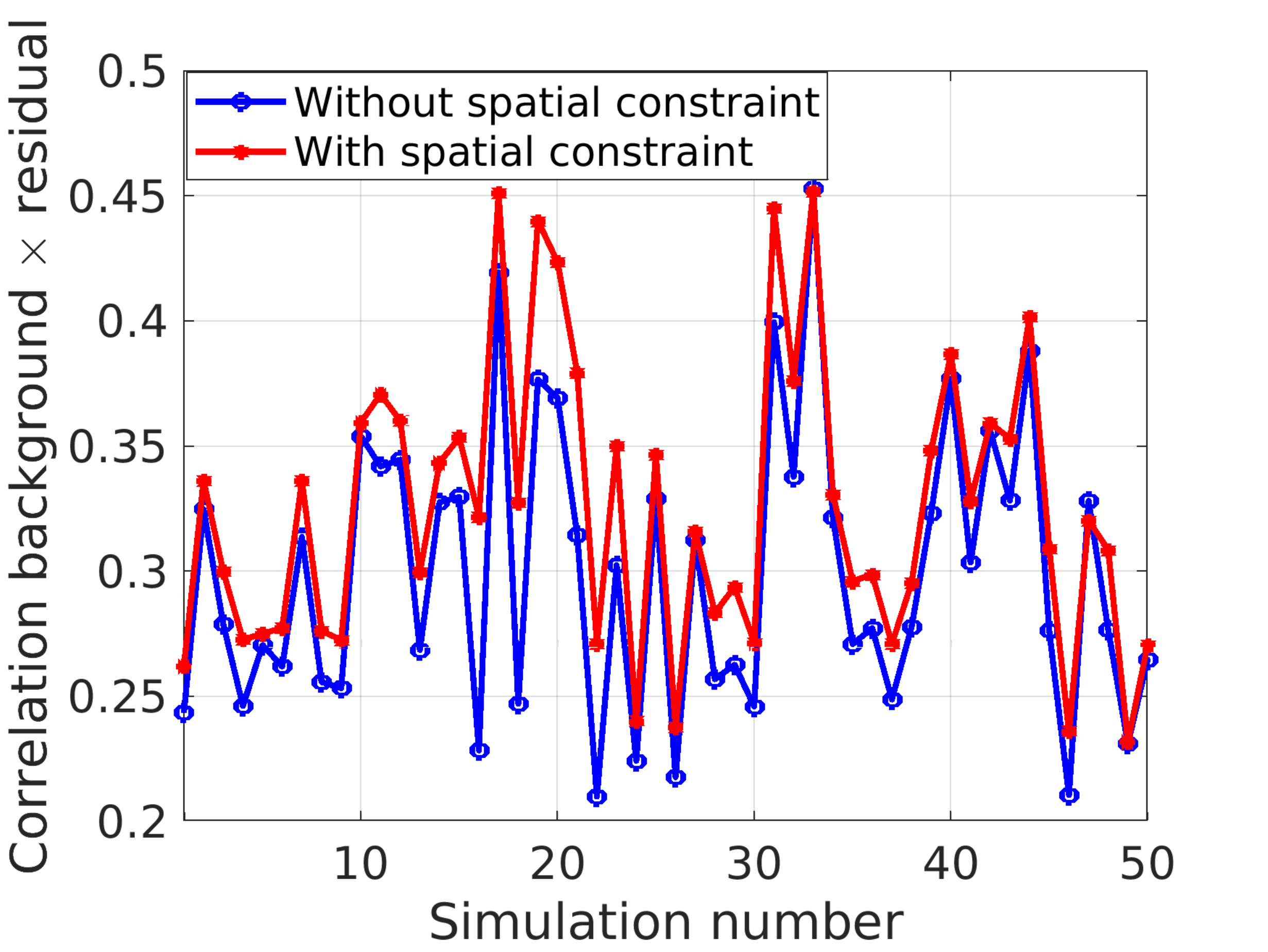,width=8.0cm}}
\vspace{0.1in}
\end{center}
\end{minipage}
  \caption{The correlation of the residual maps with the weak sky signal (in Stokes I,Q and U). With spatial constraint, we can increase the correlation from about 30\% to 34\% on average.\label{corr}}
\end{figure}

We measure the performance of calibration by measuring the suppression of the unmodeled weak sources in the sky due to calibration. As seen in Fig. \ref{simulated_maps} (b), we can simulate only the weak sky, and compare this to the residual images made after calibration. Numerically, we can cross correlate for example Fig. \ref{simulated_maps} (b) with Fig. \ref{simulated_maps} (c) or Fig. \ref{simulated_maps} (d). We show the correlation coefficient calculated this way for several simulations in Fig. \ref{corr}. Note that calibration along $K=10$ directions with only 100 sec of data is not well constrained. Nonetheless, we see from Fig. \ref{corr} that introducing spatial constraints into the problem increases the correlation of the residual with the weak sky. In other words, the spatial constraints decreases the loss of unmodeled structure in the sky.

\begin{figure}
\begin{minipage}{0.98\linewidth}
\begin{center}
\centering
\centerline{\epsfig{figure=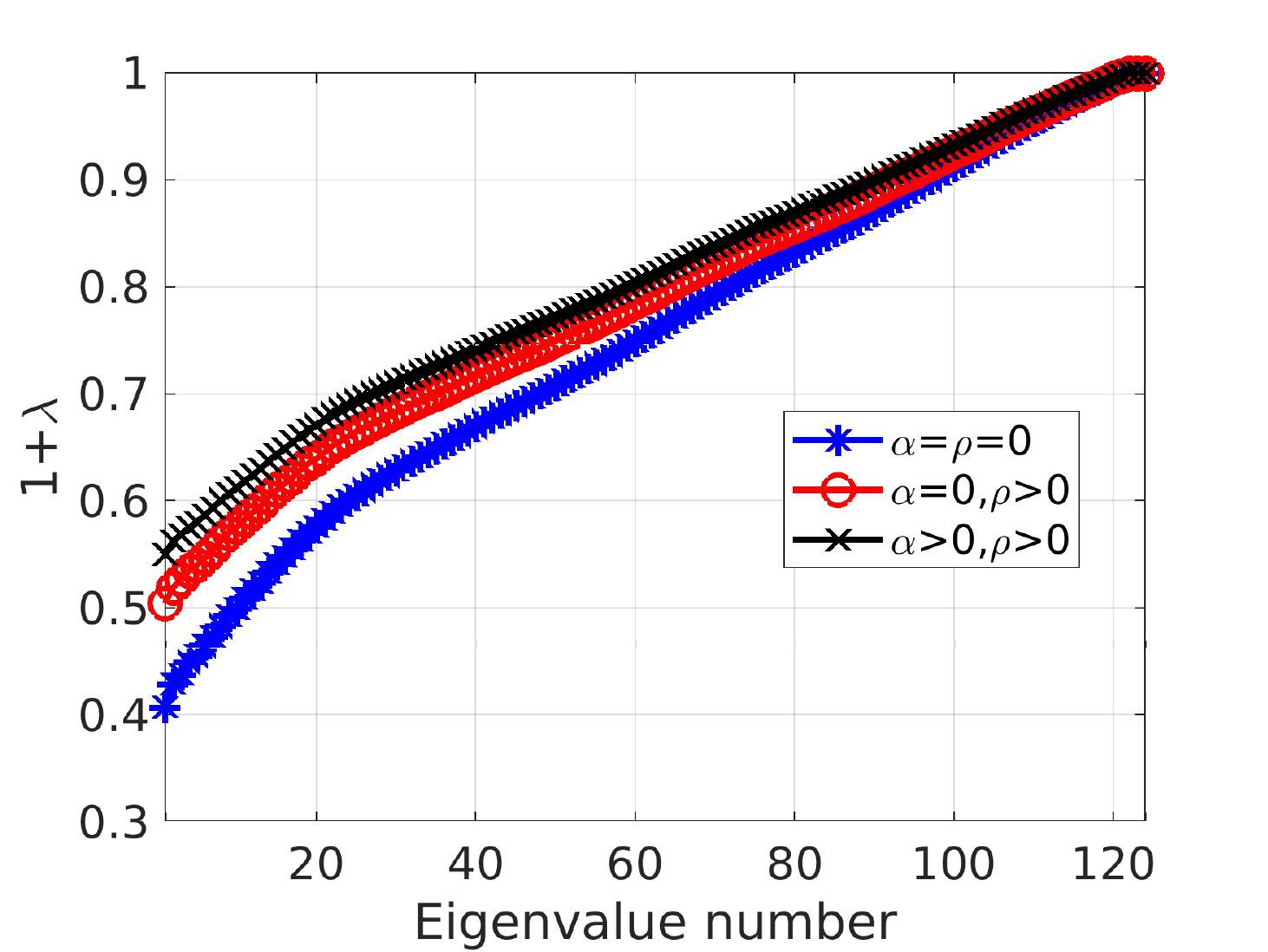,width=8.0cm}}
\vspace{0.1in}
\end{center}
\end{minipage}
  \caption{Eigenvalue plots for the influence function of calibration with (i) no regularization (ii) spectral regularization and (iii) both spectral and spatial regularization. The epigraph of the curve is lowest with both spectral and spatial regulalrization.\label{eigenvals}}
\end{figure}

We use the influence function  derived in section \ref{sec:perf} to explain the behavior in Fig. \ref{corr}. We show the plots of $1 + \lambda_i$ with $i$ in Fig. \ref{eigenvals} for calibration with no regularization ($\rho=\alpha=0$), with only spectral regularization ($\rho>0,\alpha=0$), and with both spectral and spatial regularization ($\rho>0,\alpha>0$). We see that calibration with both spatial and spectral regularization gives the curve of $1+\lambda_i$ with the lowest area above the curve (or the epigraph). In other words, the number of degrees of freedom consumed by calibration with both spectral and spatial regularization is the lowest. This also means that the loss of weak signals due to calibration is the lowest for both spectral and spatial regularized calibration, as we show in Fig. \ref{corr}. However, the improvement depends on the actual amount of deterministic variability of systematic errors with direction and the appropriate selection of regularization parameters $\rho,\alpha$.

\section{Conclusions}\label{sec:conclusions}
We have presented a method to incorporate spatial constraints to spectrally constrained direction dependent calibration. Whenever the direction dependent errors have a deterministic spatial variation, we can bootstrap a spatial model. With only a small increase in computations, we are able to improve the performance of spectrally constrained calibration. The improvement depends on the actual spatial variation of systematic errors, i.e., whether it is deterministic or stochastic. Moreover, the appropriate selection of the regularization factor $\alpha$ is also affecting the final result. Future work will focus on automating the determination of optimal regularization and modelling parameters, for example, by using reinforcement learning.

\section*{Acknowledgments}
We thank Chris Jordan for the careful review and valuable comments.

\section*{Data availability}
Ready to use software based on this work and test data are available online\footnote{http://sagecal.sourceforge.net and https://github.com/SarodYatawatta/smart-calibration}.

\bibliographystyle{mnras}
\bibliography{references}

\begin{thebibliography}{}
\makeatletter
\relax
\def\mn@urlcharsother{\let\do\@makeother \do\$\do\&\do\#\do\^\do\_\do\%\do\~}
\def\mn@doi{\begingroup\mn@urlcharsother \@ifnextchar [ {\mn@doi@}
  {\mn@doi@[]}}
\def\mn@doi@[#1]#2{\def\@tempa{#1}\ifx\@tempa\@empty \href
  {http://dx.doi.org/#2} {doi:#2}\else \href {http://dx.doi.org/#2} {#1}\fi
  \endgroup}
\def\mn@eprint#1#2{\mn@eprint@#1:#2::\@nil}
\def\mn@eprint@arXiv#1{\href {http://arxiv.org/abs/#1} {{\tt arXiv:#1}}}
\def\mn@eprint@dblp#1{\href {http://dblp.uni-trier.de/rec/bibtex/#1.xml}
  {dblp:#1}}
\def\mn@eprint@#1:#2:#3:#4\@nil{\def\@tempa {#1}\def\@tempb {#2}\def\@tempc
  {#3}\ifx \@tempc \@empty \let \@tempc \@tempb \let \@tempb \@tempa \fi \ifx
  \@tempb \@empty \def\@tempb {arXiv}\fi \@ifundefined
  {mn@eprint@\@tempb}{\@tempb:\@tempc}{\expandafter \expandafter \csname
  mn@eprint@\@tempb\endcsname \expandafter{\@tempc}}}

\bibitem[\protect\citeauthoryear{{Albert}, {Oei}, {van Weeren}, {Intema}  \&
  {R{\"o}ttgering}}{{Albert} et~al.}{2020}]{Albert2020}
{Albert} J.~G.,  {Oei} M.~S.~S.~L.,  {van Weeren} R.~J.,  {Intema} H.~T.,
  {R{\"o}ttgering} H.~J.~A.,  2020, \mn@doi [\aap]
  {10.1051/0004-6361/201935668}, \href
  {https://ui.adsabs.harvard.edu/abs/2020A&A...633A..77A} {633, A77}

\bibitem[\protect\citeauthoryear{{Arora}, {Morgan}, {Ord}, {Tingay}
  et~al.}{{Arora} et~al.}{2016}]{arora2}
{Arora} B.~S.,  {Morgan} J.,  {Ord} S.~M.,  {Tingay} S.~J.,   et~al., 2016,
  \mn@doi [\pasa] {10.1017/pasa.2016.22}, \href
  {http://adsabs.harvard.edu/abs/2016PASA...33...31A} {33, e031}

\bibitem[\protect\citeauthoryear{Beck \& Teboulle}{Beck \&
  Teboulle}{2009}]{FISTA}
Beck A.,  Teboulle M.,  2009, \mn@doi [SIAM Journal on Imaging Sciences]
  {10.1137/080716542}, 2, 183

\bibitem[\protect\citeauthoryear{{Bhatnagar}, {Cornwell}, {Golap}  \&
  {Uson}}{{Bhatnagar} et~al.}{2008}]{B2008}
{Bhatnagar} S.,  {Cornwell} T.~J.,  {Golap} K.,   {Uson} J.~M.,  2008, \mn@doi
  [\aap] {10.1051/0004-6361:20079284}, \href
  {https://ui.adsabs.harvard.edu/abs/2008A&A...487..419B} {487, 419}

\bibitem[\protect\citeauthoryear{Boyd, Parikh, Chu, Peleato  \& Eckstein}{Boyd
  et~al.}{2011}]{boyd2011}
Boyd S.,  Parikh N.,  Chu E.,  Peleato B.,   Eckstein J.,  2011, Foundations
  and Trends{\textregistered} in Machine Learning, 3, 1

\bibitem[\protect\citeauthoryear{{Brossard}, {El Korso}, {Pesavento}, {Boyer},
  {Larzabal}  \& {Wijnholds}}{{Brossard} et~al.}{2016}]{Brossard2016}
{Brossard} M.,  {El Korso} M.~N.,  {Pesavento} M.,  {Boyer} R.,  {Larzabal} P.,
    {Wijnholds} S.~J.,  2016, preprint, \href
  {http://adsabs.harvard.edu/abs/2016arXiv160902448B} {} (\mn@eprint {arXiv}
  {1609.02448})

\bibitem[\protect\citeauthoryear{{Cotton}}{{Cotton}}{2007}]{Cotton}
{Cotton} W.,  2007, Very Large Array (VLA) Scientific Memorandum 118, pp 1--12

\bibitem[\protect\citeauthoryear{{Cotton} \& {Mauch}}{{Cotton} \&
  {Mauch}}{2021}]{Cotton2021}
{Cotton} W.~D.,  {Mauch} T.,  2021, arXiv e-prints, \href
  {https://ui.adsabs.harvard.edu/abs/2021arXiv210910151C} {p. arXiv:2109.10151}

\bibitem[\protect\citeauthoryear{{Fessler} \& {Hero}}{{Fessler} \&
  {Hero}}{1994}]{Fess94}
{Fessler} J.,  {Hero} A.,  1994, IEEE Trans. on Sig. Proc., 42, 2664

\bibitem[\protect\citeauthoryear{{Hamaker}, {Bregman}  \& {Sault}}{{Hamaker}
  et~al.}{1996}]{HBS}
{Hamaker} J.~P.,  {Bregman} J.~D.,   {Sault} R.~J.,  1996, Astronomy and
  Astrophysics Supp., 117, 137

\bibitem[\protect\citeauthoryear{Hampel, Ronchetti, Rousseeuw  \&
  Stahel}{Hampel et~al.}{1986}]{Hampel86}
Hampel F.~R.,  Ronchetti E.,  Rousseeuw P.~J.,   Stahel W.~A.,  1986, Robust
  statistics: the approach based on influence functions.
New York USA:Wiley, \url {https://archive-ouverte.unige.ch/unige:23238}

\bibitem[\protect\citeauthoryear{{Intema}, {van der Tol}, {Cotton}, {Cohen},
  {van Bemmel}  \& {R{\"o}ttgering}}{{Intema} et~al.}{2009}]{Intema}
{Intema} H.~T.,  {van der Tol} S.,  {Cotton} W.~D.,  {Cohen} A.~S.,  {van
  Bemmel} I.~M.,   {R{\"o}ttgering} H.~J.~A.,  2009, \mn@doi [\aap]
  {10.1051/0004-6361/200811094}, 501, 1185

\bibitem[\protect\citeauthoryear{{Kazemi}, {Yatawatta}, {Zaroubi},
  {Labropoluos}, {de Bruyn}, {Koopmans}  \& {Noordam}}{{Kazemi}
  et~al.}{2011}]{Kaz2}
{Kazemi} S.,  {Yatawatta} S.,  {Zaroubi} S.,  {Labropoluos} P.,  {de Bruyn} A.,
   {Koopmans} L.,   {Noordam} J.,  2011, \mnras, 414, 1656

\bibitem[\protect\citeauthoryear{{Lu} \& {Yang}}{{Lu} \& {Yang}}{2015}]{Lu2015}
{Lu} Y.,  {Yang} J.,  2015, arXiv e-prints, \href
  {https://ui.adsabs.harvard.edu/abs/2015arXiv150700333L} {p. arXiv:1507.00333}

\bibitem[\protect\citeauthoryear{{McMahan}, {Moore}, {Ramage}, {Hampson}  \&
  {Ag{\"u}era y Arcas}}{{McMahan} et~al.}{2016}]{McMahan2016}
{McMahan} B.~H.,  {Moore} E.,  {Ramage} D.,  {Hampson} S.,   {Ag{\"u}era y
  Arcas} B.,  2016, arXiv e-prints, \href
  {https://ui.adsabs.harvard.edu/abs/2016arXiv160205629B} {p. arXiv:1602.05629}

\bibitem[\protect\citeauthoryear{Mitchell, Greenhill, Wayth, Sault, Lonsdale,
  Cappallo, Morales  \& Ord}{Mitchell et~al.}{2008}]{RTS}
Mitchell D.~A.,  Greenhill L.~J.,  Wayth R.~B.,  Sault R.~J.,  Lonsdale C.~J.,
  Cappallo R.~J.,  Morales M.~F.,   Ord S.~M.,  2008, \mn@doi [IEEE Journal of
  Selected Topics in Signal Processing] {10.1109/JSTSP.2008.2005327}, 2, 707

\bibitem[\protect\citeauthoryear{Ollier, Korso, Ferrari, Boyer  \&
  Larzabal}{Ollier et~al.}{2018}]{OLLIER2018}
Ollier V.,  Korso M. N.~E.,  Ferrari A.,  Boyer R.,   Larzabal P.,  2018,
  \mn@doi [Signal Processing] {https://doi.org/10.1016/j.sigpro.2018.07.024},
  153, 348

\bibitem[\protect\citeauthoryear{{Yatawatta}}{{Yatawatta}}{2015}]{DCAL}
{Yatawatta} S.,  2015, \mn@doi [\mnras] {10.1093/mnras/stv596}, 449, 4506

\bibitem[\protect\citeauthoryear{Yatawatta}{Yatawatta}{2018a}]{SAM2018}
Yatawatta S.,  2018a, in 2018 IEEE 10th Sensor Array and Multichannel Signal
  Processing Workshop (SAM). pp 485--489, \mn@doi{10.1109/SAM.2018.8448481}

\bibitem[\protect\citeauthoryear{{Yatawatta}}{{Yatawatta}}{2018b}]{DistModel}
{Yatawatta} S.,  2018b, in 2018 IEEE International Conference on Acoustics,
  Speech and Signal Processing (ICASSP). pp 3489--3493,
  \mn@doi{10.1109/ICASSP.2018.8462230}

\bibitem[\protect\citeauthoryear{Yatawatta}{Yatawatta}{2019a}]{ST2019}
Yatawatta S.,  2019a, \mn@doi [Monthly Notices of the Royal Astronomical
  Society] {10.1093/mnras/stz1222}, 486, 5646

\bibitem[\protect\citeauthoryear{{Yatawatta}}{{Yatawatta}}{2019b}]{SP2019}
{Yatawatta} S.,  2019b, \mn@doi [\mnras] {10.1093/mnras/stz1222}, 486, 5646

\bibitem[\protect\citeauthoryear{Yatawatta}{Yatawatta}{2020}]{Y2020}
Yatawatta S.,  2020, \mn@doi [Monthly Notices of the Royal Astronomical
  Society] {10.1093/mnras/staa648}, 493, 6071

\bibitem[\protect\citeauthoryear{Yatawatta \& Avruch}{Yatawatta \&
  Avruch}{2021}]{Y2021}
Yatawatta S.,  Avruch I.~M.,  2021, \mn@doi [Monthly Notices of the Royal
  Astronomical Society] {10.1093/mnras/stab1401}, 505, 2141

\bibitem[\protect\citeauthoryear{{Yatawatta}, {Diblen}, {Spreeuw}  \&
  {Koopmans}}{{Yatawatta} et~al.}{2018}]{DMUX}
{Yatawatta} S.,  {Diblen} F.,  {Spreeuw} H.,   {Koopmans} L.~V.~E.,  2018,
  \mn@doi [\mnras] {10.1093/mnras/stx3130}, \href
  {http://adsabs.harvard.edu/abs/2018MNRAS.475..708Y} {475, 708}

\bibitem[\protect\citeauthoryear{Zou \& Hastie}{Zou \&
  Hastie}{2005}]{ElasticNet}
Zou H.,  Hastie T.,  2005, \mn@doi [Journal of the Royal Statistical Society:
  Series B (Statistical Methodology)] {10.1111/j.1467-9868.2005.00503.x}, 67,
  301

\makeatother
\end{thebibliography}
\bsp
\label{lastpage}
\end{document}